\documentclass{IEEEtran}
\usepackage{cite}
\usepackage{amsmath,amssymb,amsfonts}
\usepackage{graphicx}
\usepackage{textcomp,nicefrac}
\def\BibTeX{{\rm B\kern-.05em{\sc i\kern-.025em b}\kern-.08em
T\kern-.1667em\lower.7ex\hbox{E}\kern-.125emX}}
\newcommand{\kspace}{\textit{k}-space}
\newcommand{\tOne}{$T_1$}
\newcommand{\tTwo}{$T_2$}
\begin{document}
\title{Pattern-matching Unit for Medical Applications}
\author{Orlando Leombruni, Alberto Annovi, Paola Giannetti, Nicolò Vladi Biesuz, Chiara Roda, Milène Calvetti, Marco Piendibene, Luca Peretti, Matteo Cencini, Michela Tosetti, and Guido Buonincontri
\thanks{This work has been submitted to the IEEE in October 2020 for possible publication. Copyright may be transferred without notice, after which this version may no longer be accessible.}
\thanks{Orlando Leombruni. Alberto Annovi, Paola Giannetti, and Nicolò Vladi Biesuz are with Istituto Nazionale di Fisica Nucleare, Sezione di Pisa, Largo Bruno Pontecorvo 3, 56123 Pisa, Italy (e-mail: \{orlando.leombruni, alberto.annovi, paola.giannetti\}@pi.infn.it, nicolo.vladi.biesuz@cern.ch).}
\thanks{Chiara Roda, Milène Calvetti, and Marco Piendibene are with the Department of Physics of Università di Pisa, Largo Bruno Pontecorvo 3, 56123 Pisa, Italy (e-mail: \{chiara.roda, milene.calvetti\}@cern.ch, marco.piendibene@pi.infn.it).}
\thanks{Luca Peretti is with Università di Pisa and IRCCS Fondazione Stella Maris, Centro di Ricerca IMAGO7, Viale del Tirreno 331, 56128 Calambrone (Pisa), Italy (e-mail: l.peretti@studenti.unipi.it).}
\thanks{Matteo Cencini, Guido Buonincontri, and Michela Tosetti are with IRCCS Fondazione Stella Maris, Centro di Ricerca IMAGO7, Viale del Tirreno 331, 56128 Calambrone (Pisa), Italy (e-mail: \{matteo.cencini, guido.buonincontri, michela.tosetti\}@fsm.unipi.it).}}

\maketitle

\begin{abstract}
We explore the application of concepts developed in High Energy Physics (HEP) for advanced medical data analysis.

Our study case is a problem with high social impact: clinically-feasible Magnetic Resonance Fingerprinting (MRF). MRF is a new, quantitative, imaging technique that replaces multiple qualitative Magnetic Resonance Imaging (MRI) exams with a single, reproducible measurement for increased sensitivity and efficiency. A fast acquisition is followed by a pattern matching (PM) task, where signal responses are matched to entries from a dictionary of simulated, physically-feasible responses, yielding multiple tissue parameters simultaneously. Each pixel signal response in the volume is compared through scalar products with all dictionary entries to choose the best measurement reproduction. MRF is limited by the PM processing time, which scales exponentially with the dictionary dimensionality, i.e. with the number of tissue parameters to be reconstructed. We developed for HEP a powerful, compact, embedded system, optimized for extremely fast PM. This system executes real-time tracking for online event selection in the HEP experiments, exploiting maximum parallelism and pipelining.
Track reconstruction is executed in two steps. The Associative Memory (AM) ASIC first implements a PM algorithm by recognizing track candidates at low resolution. The second step, which is implemented into FPGAs (Field Programmable Gate Arrays), refines the AM output finding the track parameters at full resolution.

We propose to use this system to perform MRF, to achieve clinically reasonable reconstruction time. This paper proposes an adaptation of the HEP system for medical imaging and shows some preliminary results.
\end{abstract}


\section{Introduction}
\label{sec:introduction}
\IEEEPARstart{P}{UMA} explores the application of concepts developed to solve big data problems in High Energy Physics (HEP) for advanced analysis of medical data within the time constraint of clinical studies. In perspective, this opens the possibility of sophisticated data analysis and fast processing in many fields of imaging \cite{b1}. The PUMA study case is a new problem with high social impact: clinically-feasible Magnetic Resonance Fingerprinting (MRF) \cite{b2}, a novel multi-parametric quantitative Magnetic Resonance Imaging technique. 

Magnetic Resonance Imaging (MRI) is a powerful medical imaging technique, being able to provide images with high spatial resolution and excellent soft tissue contrast. Currently, diagnosis is performed by mean of visual inspection from the radiologist of a qualitative image of the tissue. Quantitative MRI techniques are not routinely used in clinical exams due to their long acquisition time.

MR Fingerprinting has the potential to address this problem by combining a highly under-sampled \kspace{} encoding with a continuous variation of the acquisition parameters. The variation of the acquisition parameters throughout the experiment lead to unique responses for each tissue type, which are uniquely determined by the underlying physical properties. On the other hand, the use of a highly under-sampled \kspace{} encoding allows to achieve extremely short acquisition times (on the order of 10 seconds per slice in a 2D MRI experiment).

The \kspace{} encoding scheme is designed to achieve a noise-like behavior of the aliasing artifact arising from the under-sampling. Therefore, the aliasing noise can be filtered out by comparing the acquired tissue responses and a set of physically-feasible simulated signal evolutions, yielding multiple tissue parametric maps simultaneously. In other words, the MRF technique transforms the Nuclear Magnetic Resonance (NMR) into a huge problem of pattern recognition. The standard pattern recognition algorithm is based on scalar products performed between the acquired data and the dictionary entries to choose the entry that maximizes the scalar product. 

The inclusion of further parameters into MRF may permit to tackle different aspects of disease without significantly increasing scan time. However, one of the main current limitations in increasing the number of imaged parameters is represented by the increase in pattern matching (PM) processing time, which scales exponentially with the number of entries in the dictionary. For instance, a dictionary with 6 finely sampled relevant parameters (tissue relaxation times, fat fraction estimates \cite{b3}, blood perfusion \cite{b4}) results in a dictionary with a number of entries on the order of $3\cdot10^8$. The reconstruction of an image of size $200 \times 200 \times 200$ voxels using such a dictionary and the standard reconstruction method requires more than $2\cdot10^{15}$ scalar products, taking multiple days even on a high-end workstation \cite{b5}. This prevents the adoption of such techniques in the clinical environment. Faster processing has been reached using machine learning techniques \cite{b6, b7}. As the pattern matching step is applied independently for each image voxel, the problem can be strongly parallelized and the algorithm can be accelerated by sharing the work among multiple parallel executors.

The Associative Memory (AM) ASIC \cite{b8, b9}, in cooperation with FPGAs, has a revolutionary potential to speed up the pattern matching process, being able to execute $10^{14}$ comparison instructions per second. With respect to commercial Content Addressable Memories (CAMs) it has the key feature of searching for correlation in the data. The memory access bandwidth and number of comparisons per second has, to the best of our knowledge, no equal in commercial resources. It takes full advantage of the intrinsic parallel nature of the problem by comparing in a single clock cycle, a voxel of the image under analysis to the whole set of pre-calculated \emph{expectations}, or patterns, which can be derived by offline calculation from the dictionary entries. This approach greatly reduces the complexity of the MRF reconstruction. In this study, we explored different possibilities to use the concepts of the AM accelerator to bring this system into everyday clinical use.

\section{The Associative Memory device}
\begin{figure}[ht]
\centerline{\includegraphics[width=0.8\linewidth]{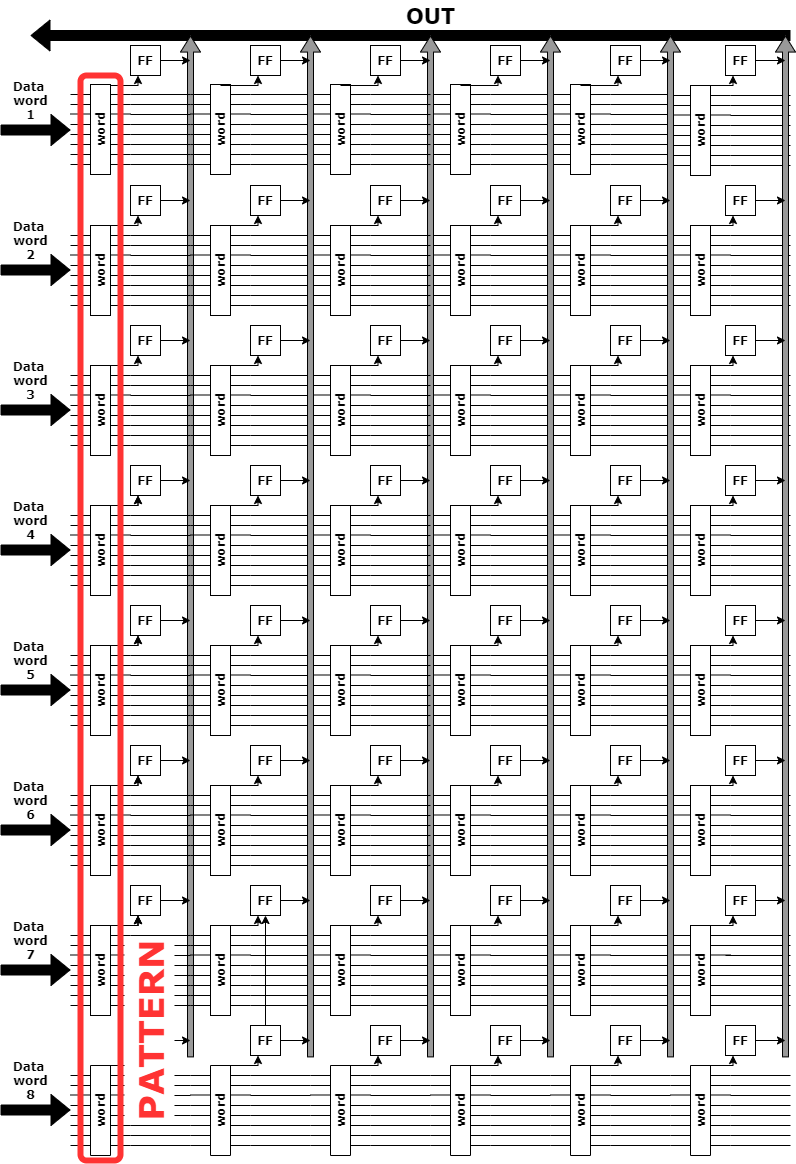}}
\caption{Architecture of an AM chip.}
\label{fig1}
\end{figure}
\begin{figure}[ht]
\centerline{\includegraphics[width=0.8\linewidth]{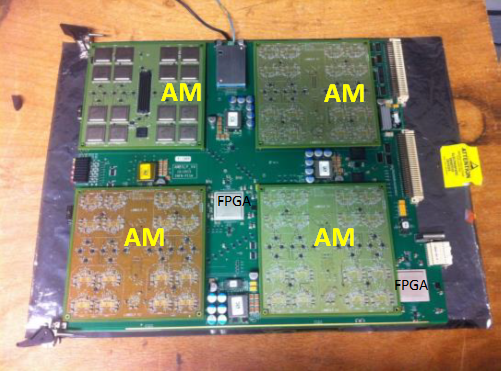}}
\caption{The AM provided of 64 AM chips and 2 large FPGAs that control the flux of data.}
\label{fig2}
\end{figure}

\figurename~\ref{fig1} shows a sketch of the architecture of the AM chip. Each pattern corresponds to a column and it is made of 8 words, each one provided of a comparator that allows an autonomous comparison with the content of the bus. The data from the MRI are provided through 8 buses in parallel. When data is introduced in the bus it is compared with all the words attached to that bus in a single clock cycle. If a word matches the input the corresponding flip flop is set to 1. A pattern is declared matched if the number of set flip flops is greater than a predefined threshold. The threshold can be set between 6 (two words of a column can miss the match) and 32, requiring the matching of one or two or four columns of words placed near in the bank.  All the matched patterns are readout and processed by an FPGA placed in the same board, like the input data are received and distributed to the AM chips by a second FPGA~\cite{b10}. \figurename~\ref{fig2} shows the board with 2 large FPGAs controlling the flux of data to and from 4 mezzanines each one assembled with 16 AM chips.

\section{Trial scenario description}\label{tsd}
We tested the AM system on a relatively easy, yet significant scenario: the MRF reconstruction of a 3D brain image using a dictionary generated from two parameters, i.e. the longitudinal and transverse relaxation times (\tOne{} and \tTwo{}) of the tissues. This scenario certainly does not require dedicated hardware, as even a general-purpose computer can perform this kind of reconstruction in an adequate timeframe. Nevertheless, it represents a widely studied problem for which there exist a number of different solving techniques, ranging from classical parallel algorithms to modern machine learning approaches, from which we can draw comparisons to our AM-based system. In this scenario \tOne{} ranges from 0 to few seconds, while \tTwo{} ranges from 0 to few tenths of a second. A dictionary generated from these parameters has about $1.5\cdot10^5$ entries, depending on the chosen range and step for both parameters.

The MR acquisition used in our trial is a 3D image where each voxel is associated to a signal sampled in multiple time points. The acquisition is therefore represented as a 4D matrix of complex numbers. Often, to reduce the computational complexity, data pertaining both the MR signal and the dictionary are projected onto a low-rank subspace determined by a Singular Value Decomposition (SVD) of the MRF dictionary~\cite{b11}. This allows for a substantial reduction of the matrix sizes, as only a small fraction of SVD coefficients is taken for each voxel.

\section{Reconstruction speed-up using the AM}\label{speedup}
The AM board stores the data to be matched (in this case, the simulated signals in the dictionary) into patterns, themselves made up by multiple 16-bit words. Since a dictionary entry is a vector of complex numbers, it must be converted into a format that is suitable to be stored into the AM banks. As the bounds of each component are known, we employed a binning technique that converts each component of the dictionary entry vector into two integer numbers, one for the real part and one for the imaginary part, such that two close dictionary components are converted into the same bin (pair of integers). Each dictionary entry is thus converted from a complex vector to a sequence of pairs of integers, and entries that show a strong degree of correlation are represented by the same sequence, i.e. the same pattern. Each pair of integers in the sequence is stored into two-word registers in the AM. Finally, one of the advantages of using the AM board to perform pattern matching is that it can be configured to allow partial matches, i.e. an input can match a pattern even if some words do not coincide. This turns the “exact match search” nature of the pattern matching into a search for correlation between the input and the dictionary entries, resulting in a more efficient algorithm, as voxels are reconstructed even when their signal does not match exactly any of the dictionary entries – without increasing the computational cost.

The procedure is carried out as follows. The dictionary is pre-processed to add simulated noise and each entry is converted into an AM pattern, which is then loaded into the board~\cite{b12}. The list of dictionary entries that are converted into the same pattern is associated to the pattern itself, for later use.
In the clinical trial, each voxel of the MR image is converted into the pattern format and fed to the system as input of the board. The AM will then perform pattern matching, which can have three outcomes: no match, (possibly multiple) partial matches, full match. In the latter two cases, the list of full resolution dictionary entries associated to the matched pattern(s) is retrieved and the FPGA will compute the dot products between those entries and the full resolution voxel currently being processed, to find the dictionary entry that better match the voxel. In case of no match, the voxel is reconstructed with the standard method by performing dot products with the entire dictionary.

In summary, binning divides the dictionary entries into “categories” of sorts, the patterns, where the entries that are binned into the same pattern show some degree of correlation (similar values for the different tissue parameters). The AM selects, for each voxel, the category (or categories) of dictionary entries which show higher degrees of correlation, and then instructs the FPGA to compute the scalar products with the entries in those categories. 
The number of categories is a function of the binning operation. Smaller bins increase the number of categories, therefore reducing the number of entries in each category and speeding up the FPGA dot product phase. In this case, entries in the same category show stronger correlation, i.e.~they present very close values for the tissue parameters. The trade-off is that the pattern matching phase efficiency gets lower as the bins become smaller, since the matching is more strict.

\section{Experimental Validation}
We tested the correctness and quality of the AM board solution using a custom simulator that allowed us to fine-tune all the reconstruction parameters, e.g.~bin size, number of SVD coefficients and more. This allowed for a rapid prototyping and validation of the solution, without having to continuously deal with the AM configuration and bank loading times.
The procedure was tested on a MR image of a brain, consisting of 200x200x200 voxels. Acquisition and reconstruction details, including dictionary generation follow Gomez et al.~\cite{b7} and are here briefly summarized.

\subsection{Acquisition} 
\begin{figure}[ht]
\centerline{\includegraphics[width=0.8\linewidth]{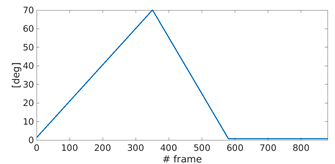}}
\caption{Variable Flip Angle scheme used for the acquisition.}
\label{fig3}
\end{figure}

A healthy human volunteer was scanned on a GE HDxt MR scanner (1.5T) using an 8ch receiver head coil, in compliance with our ethical approvals. Our technique consists of an Inversion Recovery-prepared Steady-State Free Precession (SSFP) sequence \cite{b13} using a variable Flip Angle schedule throughout the experiment (see \figurename~\ref{fig3}). \kspace{} encoding is based on a 3D spiral projection trajectory \cite{b14}. The whole acquisition is repeated with the same Flip Angle schedule; as a result, each time-point of the Flip Angle schedule is sampled with 56 interleaves of the spiral.

Echo Time (TE) and Repetition Time (TR) are kept constant during the whole experiment (TE/TR = 0.5ms/12ms). At the end of each TR, a spoiler gradient is applied to achieve a $4\pi$ dephasing along the $z$ direction.

\subsection{Reconstruction} To alleviate the impact of under-sampling artifact, data were first pre-processed by applying the so-called \kspace{} weighted view-sharing technique~\cite{b15}. Briefly, the amount of sampling of each individual frame is increased by sharing data across neighboring frames. As the spiral trajectory used to sample the \kspace{} has a radial symmetry, and is more sampled towards the center of \kspace{} than in the edges, the amount of sharing was chosen to be proportional to the sampling density. As a result, the sharing increase towards the edges of \kspace{}, while the center retains the original data. Since the center of the \kspace{} represents the contrast information of the image, this choice allows to reduce the under-sampling artifact while preserving the original shaper of the signal evolution. 
After view-sharing, \kspace{} data are projected onto a low-rank subspace as explained in Section~\ref{tsd}. Here, we retained the first 8 SVD coefficients. Then, compressed \kspace{} data were transformed to image space using a Non-Uniform Fourier transform, obtaining a set of SVD coefficient images for each individual channel of the receiver coil. First SVD coefficient images from each receiver channel were used to estimate coil sensitivity maps~\cite{b16}, and finally the different channels were combined obtaining a single set of SVD coefficients to be fed into the reconstruction algorithm. To reduce the computational burden of the inference step, the magnitude of the first SVD coefficient image was use to filter out the voxel corresponding to the background. Voxels where the magnitude of the first SVD coefficient image was lower than 0.15 times its maximum were excluded from the fit.

\subsection{Dictionary creation} A dictionary of signal evolutions was computed by simulating the tissue response to the prescribed Flip Angle schedule using the Extended Phase Graphs formalism \cite{b17}. \tOne{} and \tTwo{} values used to generate the dictionary were as follows:

\begin{itemize}
    \item \tOne{} ranges from 0.1 to 3 seconds, with a step of 0.01 seconds, for a total of 291 possible values for this parameter;
    \item \tTwo{} ranges from 0.01 to 0.6 seconds, with a step of 0.001 seconds, for a total of 591 possible values for this parameter.
\end{itemize}

In total, there are 171981 possible combinations of parameters; as such, this is also the number of entries in the dictionary. Both dictionary entries and MR signal voxels are sampled in 880 time points and then projected into a low-rank subspace determined by the SVD matrix; as said before, only the first few coefficients (in this case 8) are used.

The dictionary is pre-processed with the addition of a certain amount of noise to dictionary entries: a random quantity is added to each entry $C = \{c_1, \dots, c_8\}$. The probability density function for this random quantity is a gaussian distribution with mean 0 and standard deviation $\sigma^2$ defined as
\begin{equation}
    \sigma^2 = h\sum_{i=1}^8 c_i^2
\end{equation}
where $h$ is a constant such that $0\leq h \leq1$. The “noised” entries do not replace the original ones; rather, they are inserted as new entries in the dictionary, which therefore increases in size. The noise application process is repeated multiple times in order to increase entropy. The number of noise applications and the constant $h$ are parameters that affect the final reconstruction result; their optimal values were chosen heuristically through repeated trials on the same dictionary and MR signal. In the end, we settled on 150 repetitions done in the following way: 30 repetitions with $h = 0.01$, then 30 repetitions with $h = 0.02$ and so on until $h = 0.05$. This procedure yielded a final dictionary size of nearly 26 million entries. Please note that the addition of noise ultimately serves the purpose of increasing the number of dictionary entries in each AM category, as the combinations of the tissue parameters \tOne{} and \tTwo{} are not affected by this process (i.e. the nearly 26 million dictionary entries ultimately refer to the same 171981 different combinations of the tissue parameters in the original dictionary).

After adding the noise, the dictionary is ready to be converted into the AM bank format. Each entry’s components are binned using the technique explained in Section~\ref{speedup}, forming a pattern. Starting from the 26 million entries, only about 6000 patterns are generated. Each pattern is associated to the list of dictionary entries that generated it; in our trial, the lists are composed from a number of entries between 1 and 45000. Again, a few trials were needed in order to choose the number of bins that maximized reconstruction efficiency while keeping the error w.r.t. the standard low. We obtained the best results when choosing 15 bins for each component.

\subsection{Pattern matching} After converting the dictionary into the AM bank, the simulator converts each voxel of the signal into a pattern and then performs pattern matching with the entire bank. The output will consist of one or more patterns for each voxel, each associated to a list of dictionary entries. The simulator then computes the dot products between the voxel and all the entries found in this way, finds the maximum and associates the tissue values of the selected entry to the voxel. Voxels with no matched pattern are reconstructed in the same way as the standard technique, i.e. computing the dot products with the entire (original) dictionary.

\section{Experimental results}
\begin{figure}[htbp]
\centerline{\includegraphics[width=.9\linewidth,scale=1]{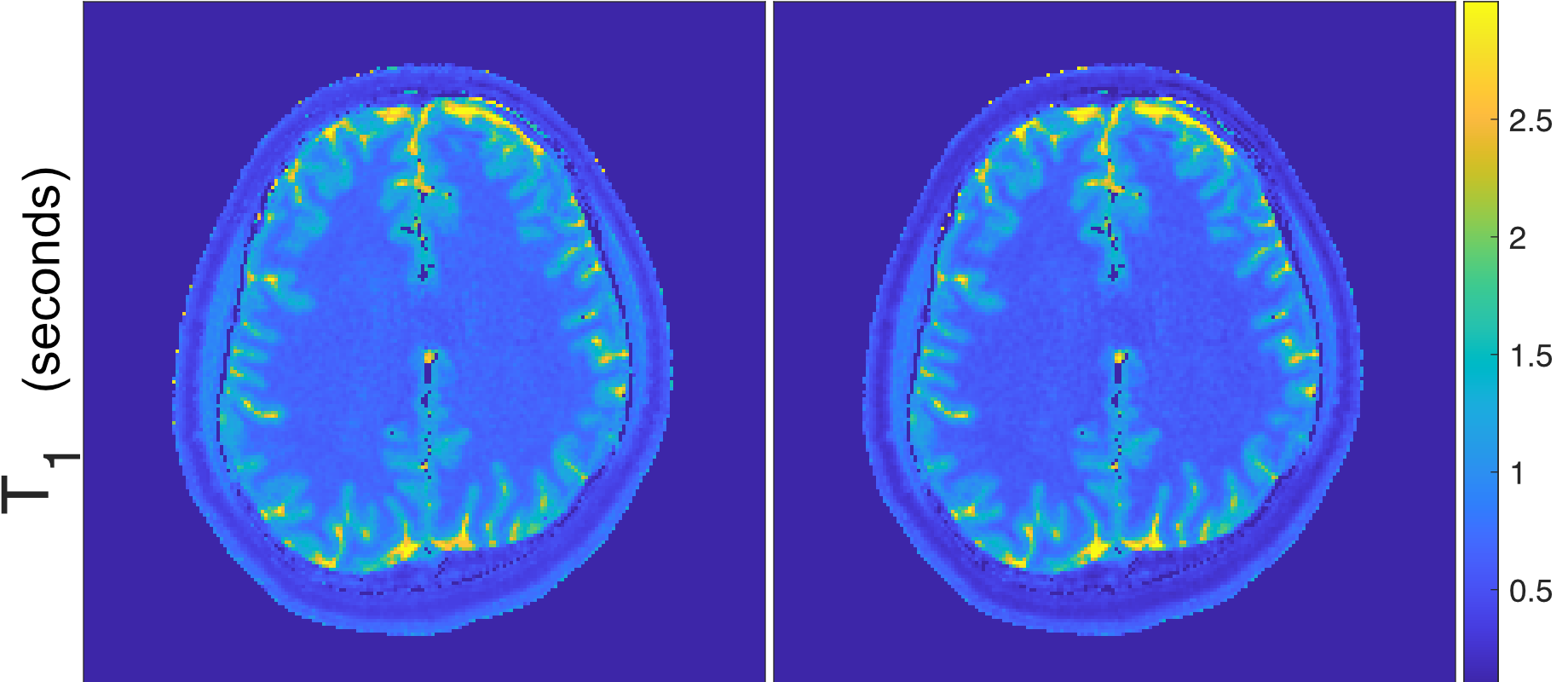}}

\centerline{\includegraphics[width=.9\linewidth,scale=1]{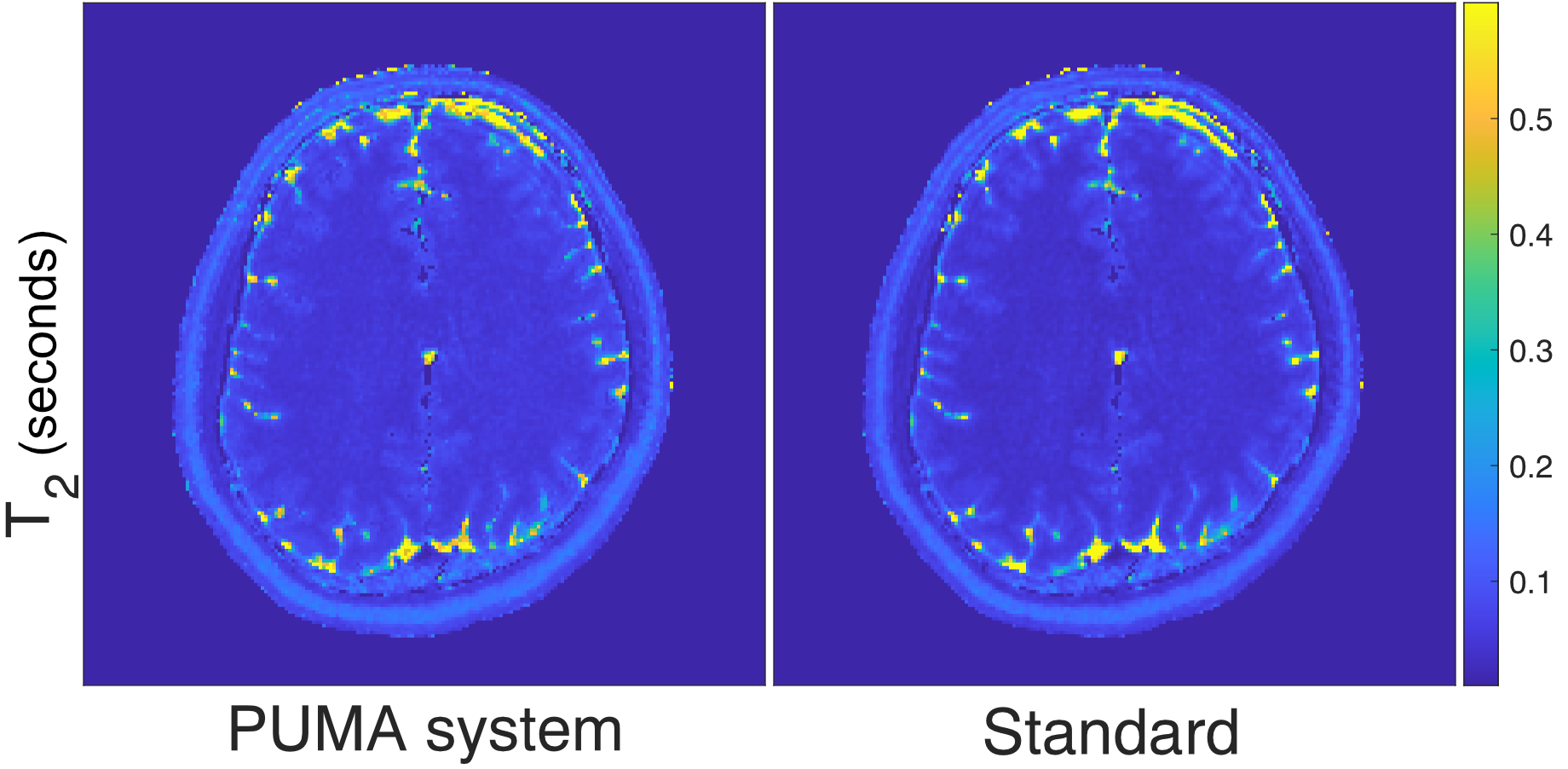}}
\caption{Comparison between reconstruction methods on the same MR acquisition for $T_1$ (top) and $T_2$ (bottom).}
\label{fig4}
\end{figure}

The results show a high degree of accuracy when comparing the image reconstructed with this method to the same image reconstructed with the “all dot products” standard technique.
\figurename~\ref{fig4} shows a slice of the image reconstructed with the PUMA method (on the left) compared with the standard method (on the right).

\begin{figure}[t]
\centerline{\includegraphics[width=.45\linewidth,scale=1]{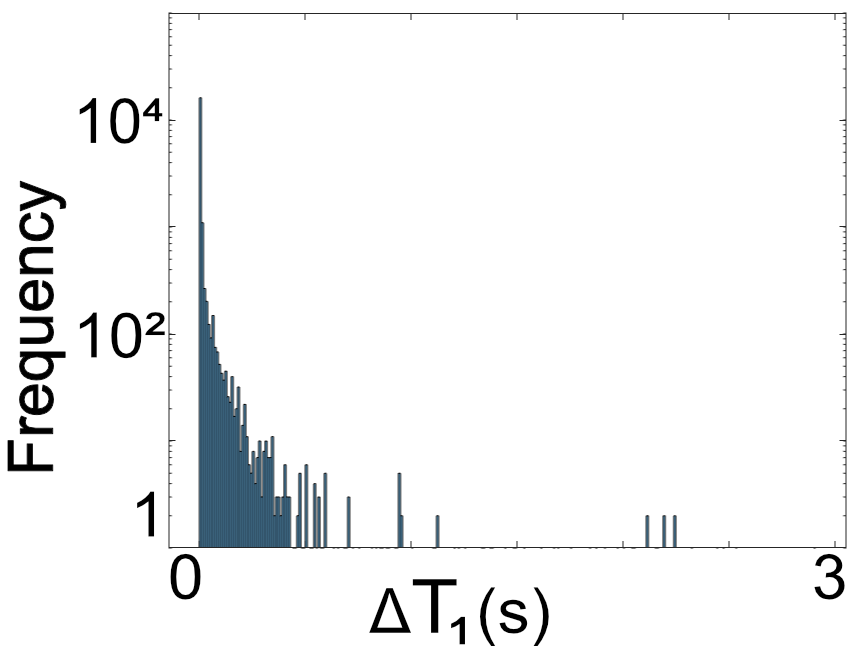} \includegraphics[width=.45\linewidth,scale=1]{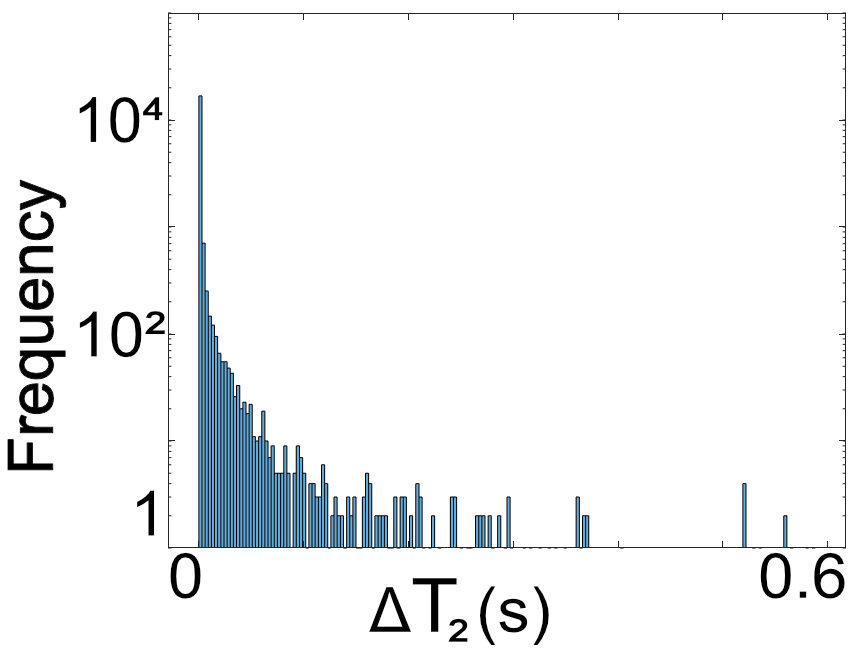}}
\caption{Frequencies (in logarithmic scale) of absolute errors for tissue parameters $T_1$ (left) and $T_2$ (right).}
\label{fig5}
\end{figure}

In particular, both mean and standard deviation for the T1 and T2 errors are very low. More than 85\% of the voxels are reconstructed with nearly-identical tissue values as the standard technique; \figurename~\ref{fig5} shows that more than $10^4$ voxels have an absolute error of less than 0.05s for \tOne{} and less than 0.005s for \tTwo{}.

\begin{figure}[t]
    \centering
    \frame{\includegraphics[width=.45\linewidth,scale=1]{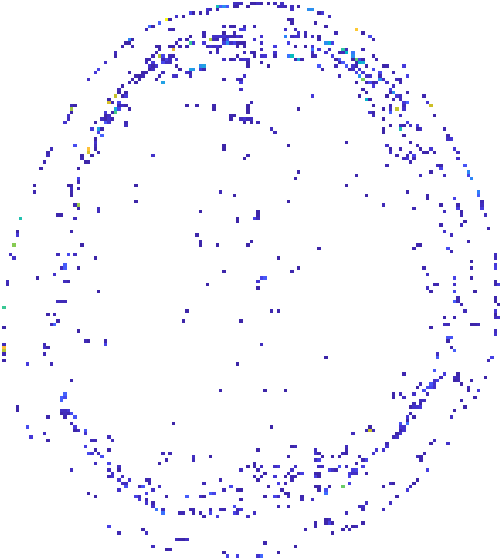}}
    \frame{\includegraphics[width=.45\linewidth,scale=1]{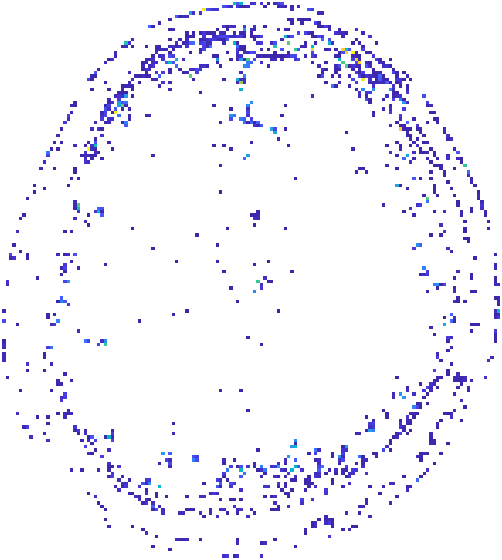}}
    \caption{Maps of voxels with absolute error higher than a certain threshold, i.e. 0.05 for \tOne{} (left), 0.005 for \tTwo{} (right)}
    \label{fig6}
\end{figure}

\figurename~\ref{fig6} shows that the absolute error is higher in perimetral tissues, while our method presents very good results (with low absolute error) in the center of the tissues.

The simulation shows that our system only needs to perform a small amount of dot products after the AM filters out uncorrelated dictionary entries. The standard reconstruction requires computing more than $6\cdot10^{11}$ dot products, whereas our system manages to achieve similar results with about $1.5\cdot10^{10}$ dot product computations, i.e. a reduction of a factor 40. Since we expect the number of dictionary entries per pattern to grow much more slowly than the dictionary size when increasing the number and grain of tissue parameters, we also expect this reduction factor to increase dramatically in those scenarios.

\section{Conclusions}
The AM and FPGA system shows very promising results for use in the MRF scenario, maintaining a high degree of accuracy in the image reconstruction while sensibly reducing the number of needed comparisons. The next step of the project will be the use of this technology for very large dictionaries produced by an higher number of parameters.

\appendices

\section*{Acknowledgement}
The results here presented have been developed in the framework of the 18HLT05 QUIERO Project. This project has received funding from the EMPIR Programme co-financed by the Participating States and from the European Union’s Horizon 2020 Research and Innovation Programme. Moreover, this work was supported in part by the Italian Ministry of Health, grant GR-2016-02361693 and RC 2017-20.

\end{document}